\documentclass[twocolumn,amsmath,floatfix,prb,aps]{revtex4}

\usepackage[utf8x]{inputenc}
\usepackage{graphicx}
\usepackage{amsmath, amsthm, amssymb}  

\usepackage{dcolumn}
\usepackage{bm}
\usepackage{color}
\usepackage{bbold}
\usepackage{bbm}

\usepackage{mathtools}        
\usepackage{braket}           
\usepackage{dsfont}           

\usepackage{float}                
\usepackage{hyperref}             
\hypersetup{pdfpagemode=UseNone}  
\hypersetup{colorlinks, citecolor=green, filecolor=black, linkcolor=blue, urlcolor=blue} 

\renewcommand{\vec}[1]{\mathbf{#1}}

\begin{document}
\DeclareGraphicsRule{*}{png}{*}{}


\author{M. Fleischmann$^{1}$}
\author{R. Gupta$^{1}$}
\author{D. Weckbecker$^{1}$}
\author{W. Landgraf$^{1}$}
\author{O. Pankratov$^{1}$}
\author{V. Meded$^2$}
\author{S. Shallcross$^{1}$}
\email{sam.shallcross@fau.de}
\affiliation{1 Lehrstuhl f\"ur Theoretische Festk\"orperphysik, Staudtstr. 7-B2, 91058 Erlangen, Germany.}
\affiliation{2 Karlsruhe Institute of Technology, Institute of Nanotechnology, Hermann-von-Helmholtz-Platz 1, 76344 Eggenstein-Leopoldshafen, Germany.}


\title{Moir\'e edge states in twisted graphene nanoribbons}
\date{\today}

\begin{abstract}

The edge physics of graphene based systems is well known to be highly sensitive to the atomic structure at the boundary, with localized zero mode edge states found only on the zigzag type termination of the lattice. Here we demonstrate that the graphene twist bilayer supports an additional class of edge states, that (i) are found for all edge geometries and thus are robust against edge roughness, (ii) occur at energies coinciding with twist induced van Hove singularities in the bulk and (iii) possess an electron density strongly modulated by the moir\'e lattice. Interestingly, these ``moir\'e edge states'' exist only for certain lattice commensurations and thus the edge physics of the twist bilayer is, in dramatic contrast to that of the bulk, not uniquely determined by the twist angle.

\end{abstract}

\maketitle

\section{Introduction}

For both single layer\cite{Brey2006a,Kobayashi2008,Gan2010} and Bernal stacked multilayer graphene\cite{Castro2008,Lima2009,Feng2009,Cortijo2010} localized zero energy boundary states are found on the zigzag terminated edges, but not for any other edge geometry. While such edge states are predicted to host a wealth of interesting physical phenomena -- including spin filtering and spin confinement\cite{son06,top08,wim08} - their evident sensitivity to edge roughness has complicated the experimental realization the experimental realization of many of these interesting effects\cite{wang16}.

The graphene twist bilayer - two mutually rotated layers of graphene - is a material with a significantly richer electronic structure than either graphene or Bernal stacked few layer graphenes. The twist bilayer\cite{hass08,shall08,shall08a,Li2009b,shall10,Bistritzer2011a,San-Jose2012,San-Jose2013,lan13,ray16} exhibits graphene like behaviour in a weak interlayer coupling large angle regime and, in contrast, charge confinement and a rich Fermiology\cite{shall13,sbo15,shall16} in a strong coupling small angle regime. This rich bulk physics might be expected to lead to correspondingly interesting edge physics, although to date this aspect of the twist bilayer has received much less attention.

In this work we consider the edge physics of the twist bilayer finding two distinct types of boundary states. The first of these are found at energies close to the Dirac point and exclusively on zigzag edges. These are essentially a modified version of the zero mode edge state found in single layer and AB stacked bilayer graphene. The second type are qualitatively different and (i) occur not at zero energy but at energies coinciding with the well known twist induced van Hove singularities\cite{yan14,hav14,yan12,oh12,brihuega12,lu11,li10,Li2009b} in the bulk electronic spectrum and, (ii), are completely insensitive to the edge geometry, being found on armchair, zigzag, as well as rough edges.

While insensitive to edge roughness, we find these ``moir\'e edge states'' are guaranteed to exist only for a certain class of lattice commensurations\cite{jia13}. This, as we show, is related to the fact that while the bulk physics of the twist bilayer is governed by a purely angle dependent Jones zone, and not the commensuration dependent Brillouin zone\cite{shall13}, this is not true in the presence of an edge. The edge physics of the twist bilayer is, therefore, sensitive to the particular twist bilayer commensuration in a way not found in the bulk electronic structure.

\section{Structure and computational method}
\label{meth}

\begin{figure*}[ht]
  \includegraphics[width=0.95\linewidth]{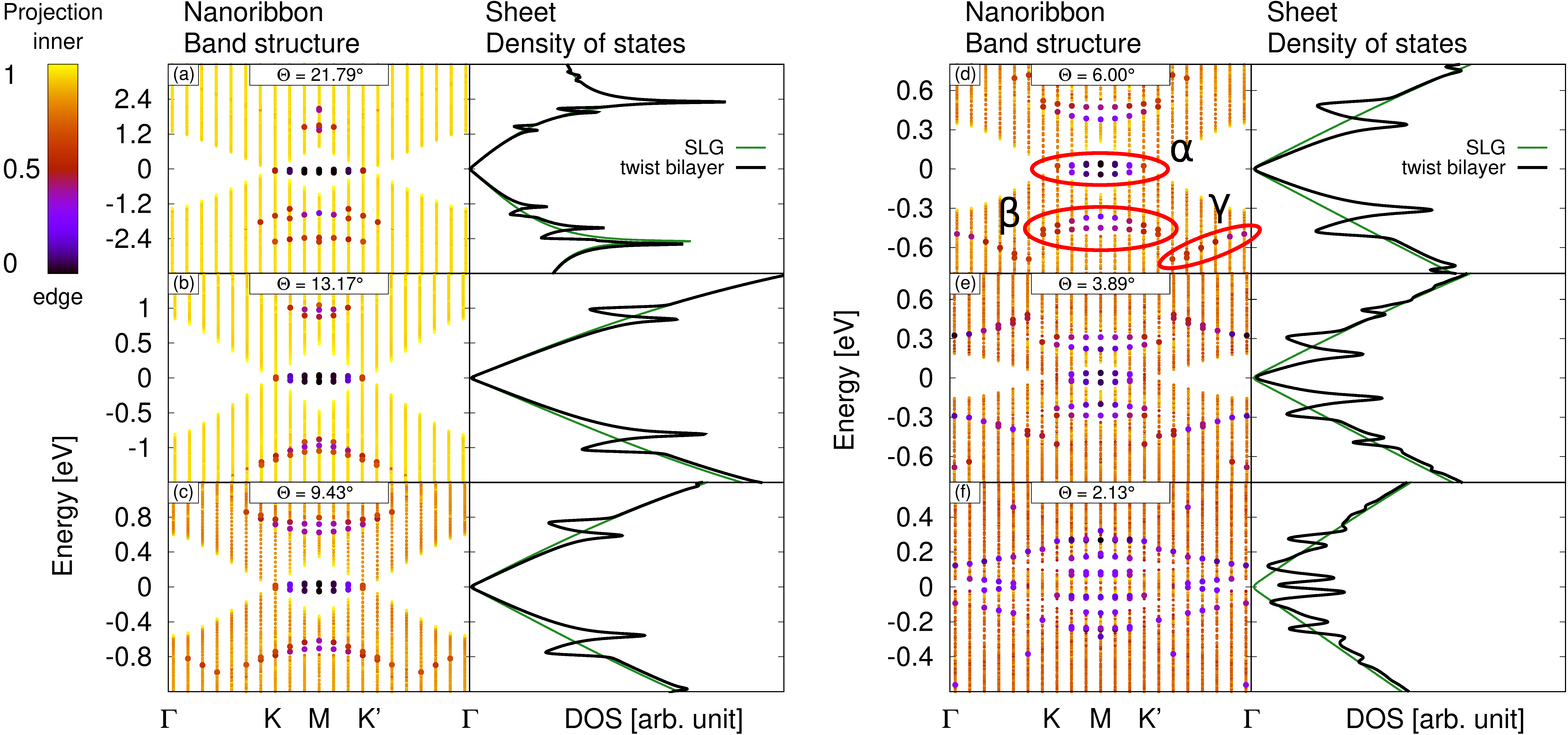}
  \caption{(Colour online.) Left column: Band structures of twist bilayer nanoribbons with $2.13^\circ < \theta < 21.79^\circ$. The colour of each eigenvalue indicates the projection of the charge density of the associated eigenstate onto the inner region of the nanoribbon (any atom within a 2~nm strip parallel and adjacent to the edge of the nanoribbon is considered to be in the ``edge region'', with the remainder of the bilayer denoted the ``inner region''). If this projection onto the inner region falls below $0.7$ the corresponding eigenvalue is highlighted by a point of larger size. The labels in panel (d) denote the three distinct types of edge states that can be seen and that are discussed in the text: zero mode edge states ($\alpha$), moir\'e edge states ($\beta$), and high energy edge states ($\gamma$). Right column: The DOS of the corresponding infinite twist sheet (black) and of single layer graphene (green). Note that the labeling of the reciprocal space axis in the band plots indicates the projection of the high symmetry points of the underlying  Brillouin zone of the twist bilayer onto the one dimensional Brillouin zone of the nanoribbon. For clarity we have rescaled the reciprocal space axis of each plot such that these projected high symmetry points coincide in each case.
}
  \label{bnd}
\end{figure*}

\emph{Construction of the nanoribbon supercells}: The problem of finding a commensuration cell of the twist bilayer leads to a discrete set of unit cells uniquely labeled by two co-prime integers, which we will refer to as $p$ and $q$ following the notation of Refs.~\onlinecite{shall10,shall13,shall16}. As shown in Ref.~\onlinecite{shall10} commensurate unit cells exist at angles given by

\begin{equation} \label{eq:dioth}
\theta = \cos^{-1}\left(\frac{3(q/p)^2-1}{3(q/p)^2+1}\right)
\end{equation}
which depends only on the ratio of $q/p$, with the number of atoms in the commensuration cell equal to

\begin{equation} \label{eq:dioN}
N = \frac{3}{\delta}\frac{1}{\gamma^2} (3q^2+p^2),
\end{equation}
where $\gamma = \text{gcd}(3q+p,3q-p)$ and $\delta = 3/\text{gcd}(p,3)$. These latter parameters label symmetry classes of the twist bilayer and, as we will show, can be used to predict which commensurations will support moir\'e edge states. The parameter $\delta$ is related to the sub-lattice exchange even (SE) and odd (SO) classes introduced by Mele\cite{mele10}, but which of the two cases $\delta = 1$ and $\delta = 3$ correspond to SE/SO depends additionally on the initial stacking of the untwisted bilayer.

For a nanoribbon geometry, these unit cells evidently define the edge structure which, as might be expected, is always a low index facet. For $p=1$ nanoribbons the edge comprises one zigzag segment per nanoribbon cell edge, with the remainder of the edge having the armchair type termination.  Nanoribbons created from cells having $p>1$ generally possess a higher ratio of zigzag to armchair edge termination. We will also consider the impact of edge roughness, which we model by randomly removing atoms from the ideal edge while ensuring that only double bonded atoms remain.

\emph{Computational method}: The large system sizes that small angle twist bilayer nanoribbons inevitably entail -- our nanoribbons are typically $\sim30$nm in width with unit cells containing up to 40,000 carbon atoms -- necessitates the use of a semi-empirical tight-binding method for the electronic structure. As we are interested only in the low energy electronic structure, we employ the Lanczos method for diagonalization. We will use the tight-binding method of Ref.~\onlinecite{lan13} which was deployed in that work for the study of twist bilayer flakes; it consists of the environment dependent method of Tang \emph{et al.}\cite{tan96}, but re-parameterized by performing a least squares fit to the high symmetry eigenvalues from a number of small unit cell few layer graphene systems generated \emph{ab initio}; for details we refer the reader to Ref.~\onlinecite{lan13}.

\section{Zero mode and moir\'e edge states}
\label{EDGE}

We first investigate the electronic structure of nanoribbons arising from $p=1$ bulk commensurations; these have one zigzag segment per edge unit cell. Our nanoribbons have a width of $\sim33$~nm and thus finite size effects play no significant role. In Fig.~\ref{bnd} are shown the band structures of six different twist bilayer nanoribbons with twist angles of $\theta = 21.79^\circ$, $13.17^\circ$, $9.43^\circ$, $6.01^\circ$, $3.89^\circ$, and $2.13^\circ$; this includes both the weak coupling (large angle) and strong coupling (small angle) limits of the twist bilayer. The colour coding of the eigenvalues refers to the projection of the corresponding state onto the edge region, defined by regions parallel and adjacent to the edges of the nanoribbon of width $\Delta L = 2$~nm. Those eigenvalues that correspond to edge states, defined as a projection of more than $0.3$ in the edge regions, are indicated by a point of larger size.

\begin{figure}[t!]
  \hspace{-1cm}
  \includegraphics[width=0.8\linewidth]{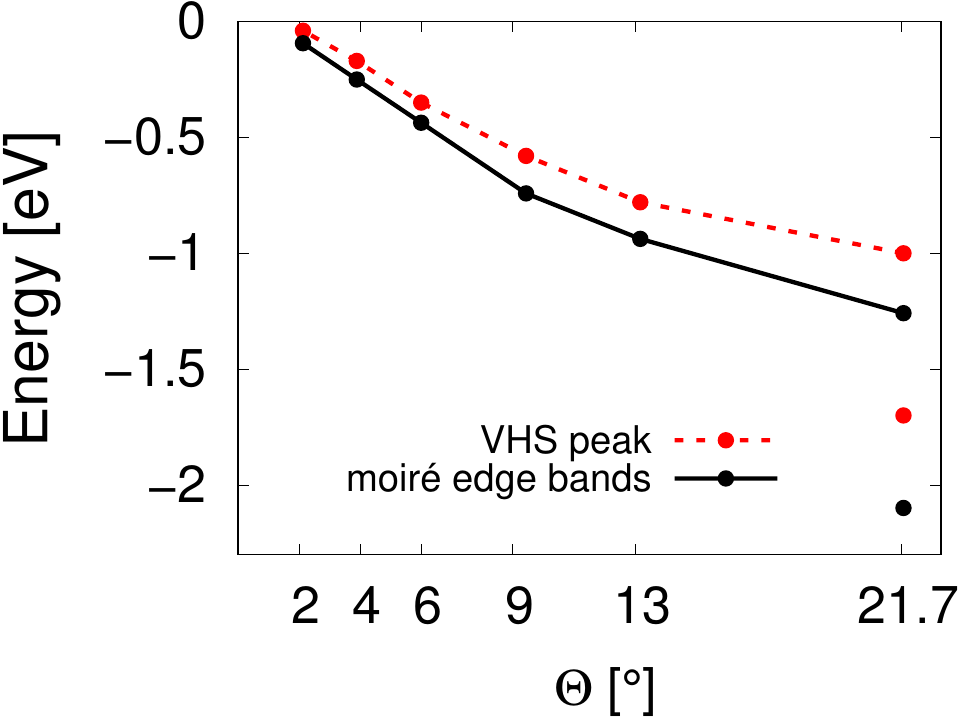}
  \caption{(Colour online.) Energy correlation between the moir\'e edge band and the low energy van Hove singularities (vHS) in the twist bilayer density of states. The full line represents the average energy of all states in the corresponding moir\'e edge band, while the dashed line represents the energy of the vHS in the twist bilayer density of state.}
  \label{correl}
\end{figure}

At each twist angle three types of edge bands can be discerned, labeled by $\alpha$, $\beta$, and $\gamma$ in Fig.~\ref{bnd}(d). The $\alpha$ edge band is always pinned near zero energy and thus represents a modified version of the zero mode edge state found in both single layer and Bernal stacked bilayer graphene. Just as for these simpler systems the zero mode band exists in the gap between the two Dirac points at $K$ and $K'$, although for the $\theta = 2.13^\circ$ nanoribbon it is instead found in the gap that extends from the $\Gamma$ point towards the $K$ and $K'$ points. This change coincides with the closing of the low energy gap between $K$ and $K'$. It may be noted that while for angles $\theta > 2.13^\circ$ the zero mode is largely independent of twist angle, there is some increase in the dispersion of the edge band as the twist angle is reduced, as well as a decrease in localization at the edge -- features already observed for the edge state in twist bilayer flakes\cite{lan13}.

\begin{figure}[t!]
  \includegraphics[width=0.90\linewidth]{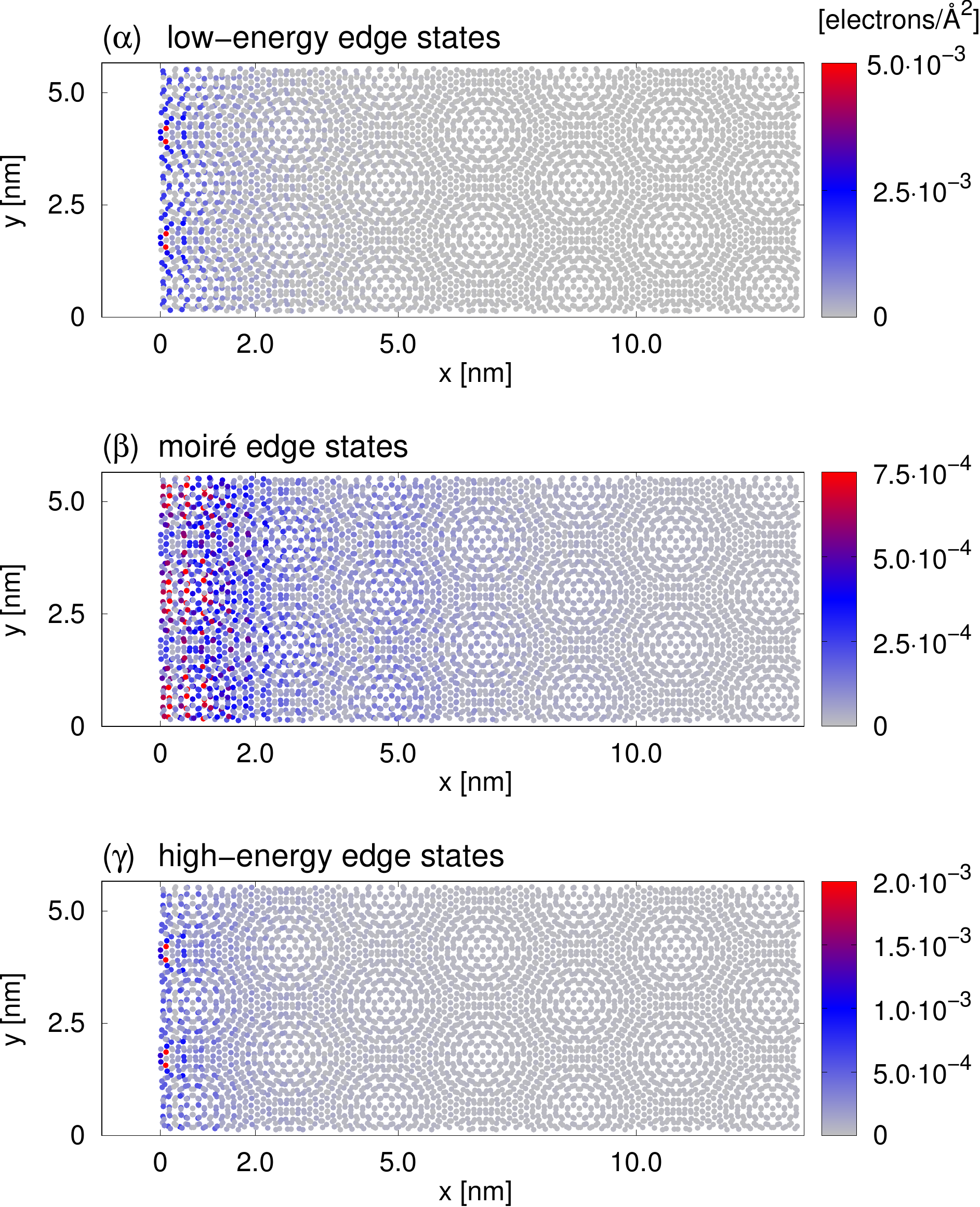}
  \caption{(Colour online.) Charge density $\rho(\vec{r})$ summed over all edge states within the highlighted regions indicated by Greek letters in panel (d) of Fig.~ \ref{bnd}: the zero mode edge states $(\alpha)$; the moir\'e edge states $(\beta)$; the high-energy edge states $(\gamma)$. The charge density distribution of the moir\'e edge states is seen to extend significantly further into the bulk of the nanoribbon than the other two types of edge states, and to show a pronounced charge localization observed on the AA patches of the moir\'e. The high density regions of the $\alpha$ and $\gamma$ edge states correspond to the zigzag portions of the nanoribbon edge.}
  \label{rho}
\end{figure}

In contrast, the moir\'e edge states, labeled by $\beta$ in Fig.~\ref{bnd}(d), are evidently closely correlated with van Hove singularities (vHS) in the bulk density of states. These edge states are found in the energy gap that arises due to the intersection and consequent level repulsion of the two Dirac cones from each layer, which both creates the vHS as well as the associated local (in momentum) energy gap; this occurs at the $M$ point in the bulk Brillouin zone, and we will refer to this as the $M$ point gap. The energetic correlation between the bulk vHS and moir\'e edge states is shown quantitatively in Fig.~\ref{correl}, where the average energy of the moir\'e edge state band is plotted along with the energy of the bulk spectrum vHS. Note that as the energy of the bulk vHS tends to zero as $\theta\to 0$, the moir\'e and zero mode edge states will, eventually, merge into one hybrid edge band at very small twist angles. Computational resources prevent us from exploring this very small angle regime here.

To further characterize these edge bands we now examine the charge density of the corresponding edge eigenstates, and for this purpose the system with $\theta = 6.00^\circ$ is chosen as a representative example. In Fig.~\ref{rho} we plot the charge density $\rho(\vec{r})$ summed over all edge states that fall within the highlighted zones shown in Fig.~\ref{bnd}{g}. Group $(\alpha)$, the zero mode edge-states, are strongly localized on the zigzag step of the otherwise armchair terminated ribbon edge (see Fig.~\ref{rho}a). This confirms that these edge states are very similar to the zero mode edge states found in single layer and Bernal stacked bilayer graphene.

In contrast, group $(\beta)$, the moir\'e edge states, are found on both zigzag and armchair terminated regions of the edge, and have a significant charge density in a region of $\approx2~\mathrm{nm}$ from the edge of the nanoribbon. Interestingly, one can clearly observe that this charge density is strongly modulated by the moir\'e, with the maxima of the charge density located on the AA spots. This demonstrates the strong relation between the moir\'e edge states and the moir\'e structure, and marks them out as quite different from the zero mode edge states. Note that the qualitative form of the integrated charge density presented here is also visible if individual states of group $(\beta)$ are plotted.

Finally, the last set of states, group $(\gamma)$, show charge density features very close to those of the zero mode. While they are energetically connected to the van Hove singularities of the twist bilayer band structure, the charge density modulation is, as for the zero mode, tied to the zigzag edge atoms and not to the moir\'e geometry.

\begin{figure}[t!]
  \hspace{-1cm}
  \includegraphics[width=0.8\linewidth]{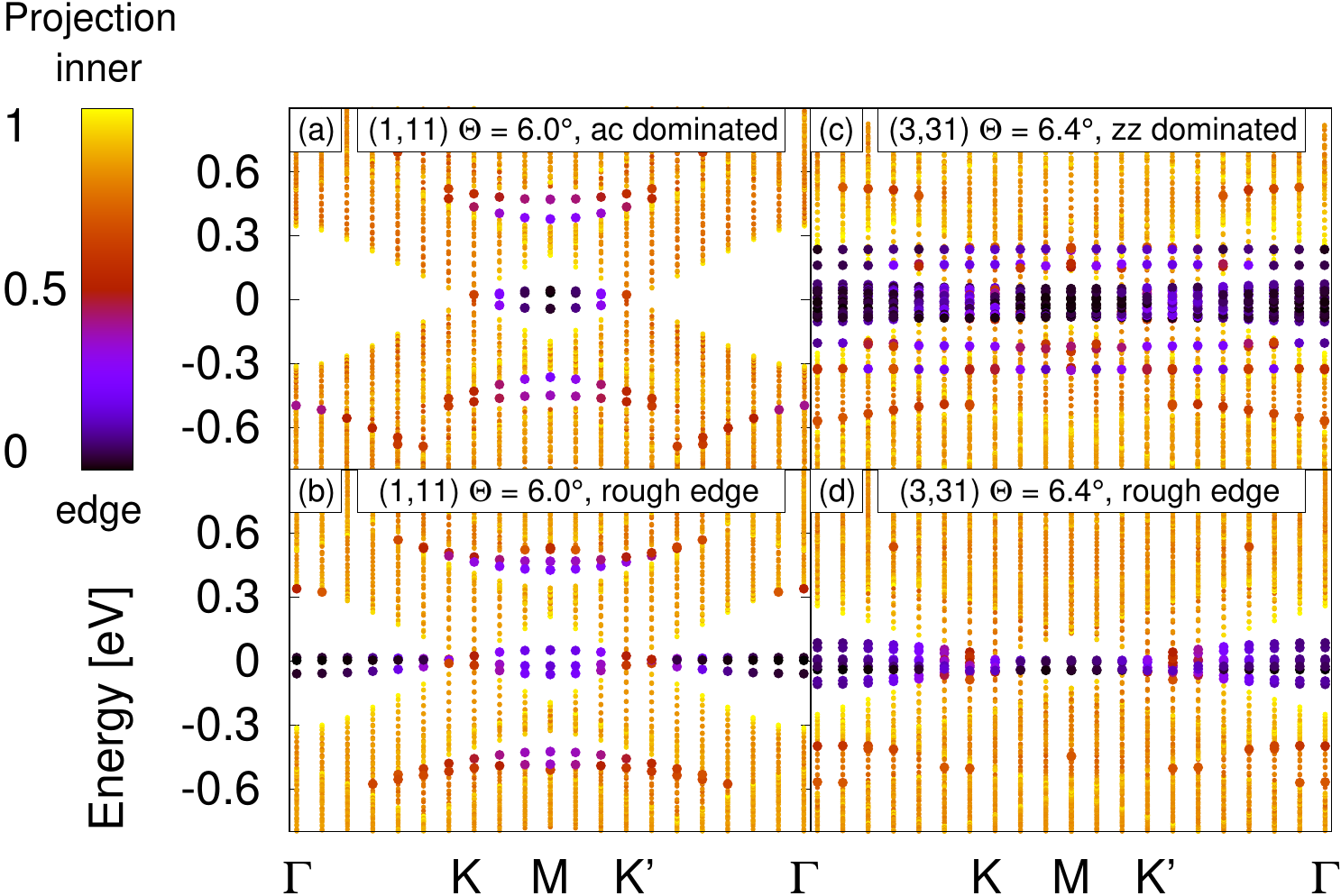}
  \caption{(Colour online.) \emph{Impact of roughness and commensuration on the moir\'e edge state}. Shown in panels (a) and (b) are, respectively, the band structure for an ideal and rough edge termination of the $(p,q) = (1,11)$ ($\theta = 6.01^\circ$) twist nanoribbon. Note that while the zero mode changes qualitatively (due to the increase in zigzag termination in the rough edge), the moir\'e edge state is largely unchanged. Panels (c) and (d) exhibit the band structure for, respectively, ideal and rough terminations of a $(p,q)=(1,10)$ commensuration (twist angle $\theta = 6.6^\circ$). While this twist angle is very similar to that of the nanoribbon band structure displayed in panels (a) and (b), for this commensuration the moir\'e edge state has vanished, a fact that can be related to the closing of the $M$ point gap. Note that the $(p,q)=(1,10)$ commensuration has zigzag dominated edges, hence the predominant zero mode found in this case.}
  \label{rbnd}
\end{figure}

\begin{figure}[t!]
  \hspace{-1cm}
  \includegraphics[width=0.8\linewidth]{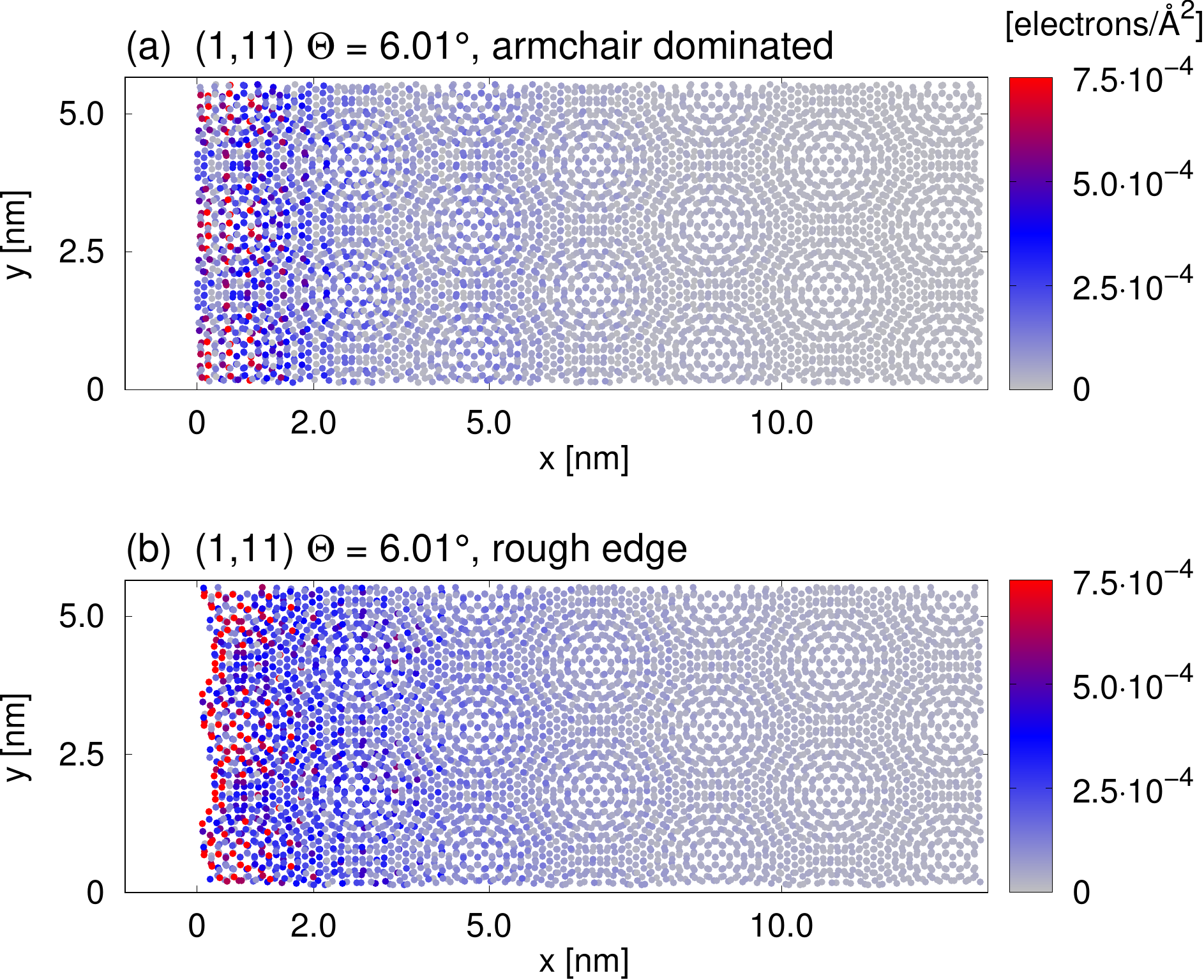}
  \caption{(Colour online.) \emph{Impact of roughness on the electron density of the moir\'e edge state}. Shown are the electron densities integrated over the edge bands for the moir\'e edge states shown in panels (a) and (b) of Fig.~\ref{rbnd}. Note that the edge state density is qualitatively unchanged by the presence of edge roughness.}
  \label{rrho}
\end{figure}

\section{Robustness of the moir\'e edge state}

We now consider the general conditions under which the moir\'e edge state can be found. As might be suspected from the fact that the charge density is localized on both zigzag and armchair terminated regions of the edge (see Fig.~\ref{rho}b), it turns out to be robust against edge roughness but, as we will show, depends in a subtle way on details of the twist bilayer commensuration. To demonstrate roughness against edge roughness we display, for both rough and ideal edges, the band structure (Fig.~\ref{rbnd}a,b) as well as the edge state charge density (Fig.~\ref{rrho}a,b) of the $(p,q) = (1,11)$ ($\theta = 6.01^\circ$) twist nanoribbon. In panel (a) of each of these figures is shown the ideal edge with panel (b) displaying -- as is readily apparent upon contrasting Figs.~\ref{rrho}(a,b) -- the same system but with significantly increased edge roughness. In both band structure as well as electron density there is no qualitative change to the moir\'e edge state upon imposition of edge roughness. In dramatic contrast, however, the nature of the zero mode changes qualitatively upon edge roughening, with this edge state now extending throughout the BZ (see Fig.~\ref{rbnd}b). This simply reflects the increase in zigzag terminated regions in the rough edge as compared to the ideal edge.

Having established that the moir\'e edge state is robust to edge roughness, we now examine how it is impacted by the choice of bulk twist bilayer commensuration. Thus far we have only considered the $p=1$, $q\in \mathbb{Z}_{\text{odd}}$ commensurations for which the unit cell vectors are equal in magnitude to the moir\'e length $D=a/(2\sin\frac{\theta}{2})$ ($a$ is the lattice parameter of graphene). These have twist angles given by $\theta_0 = \cos^{-1}\frac{3q^2-1}{3q^2+1}$, and evidently represent the minimal unit cell area close to a given $\theta_0$. There exist, however, an infinite set of larger unit cell commensurations within the angle range $\theta_0 - \epsilon < \theta < \theta_0 + \epsilon$ for any finite $\epsilon$, and a natural question is then how the edge physics depends on this choice of commensuration.

\begin{figure}[t!]
    \centering
    \includegraphics[width=0.8\linewidth]{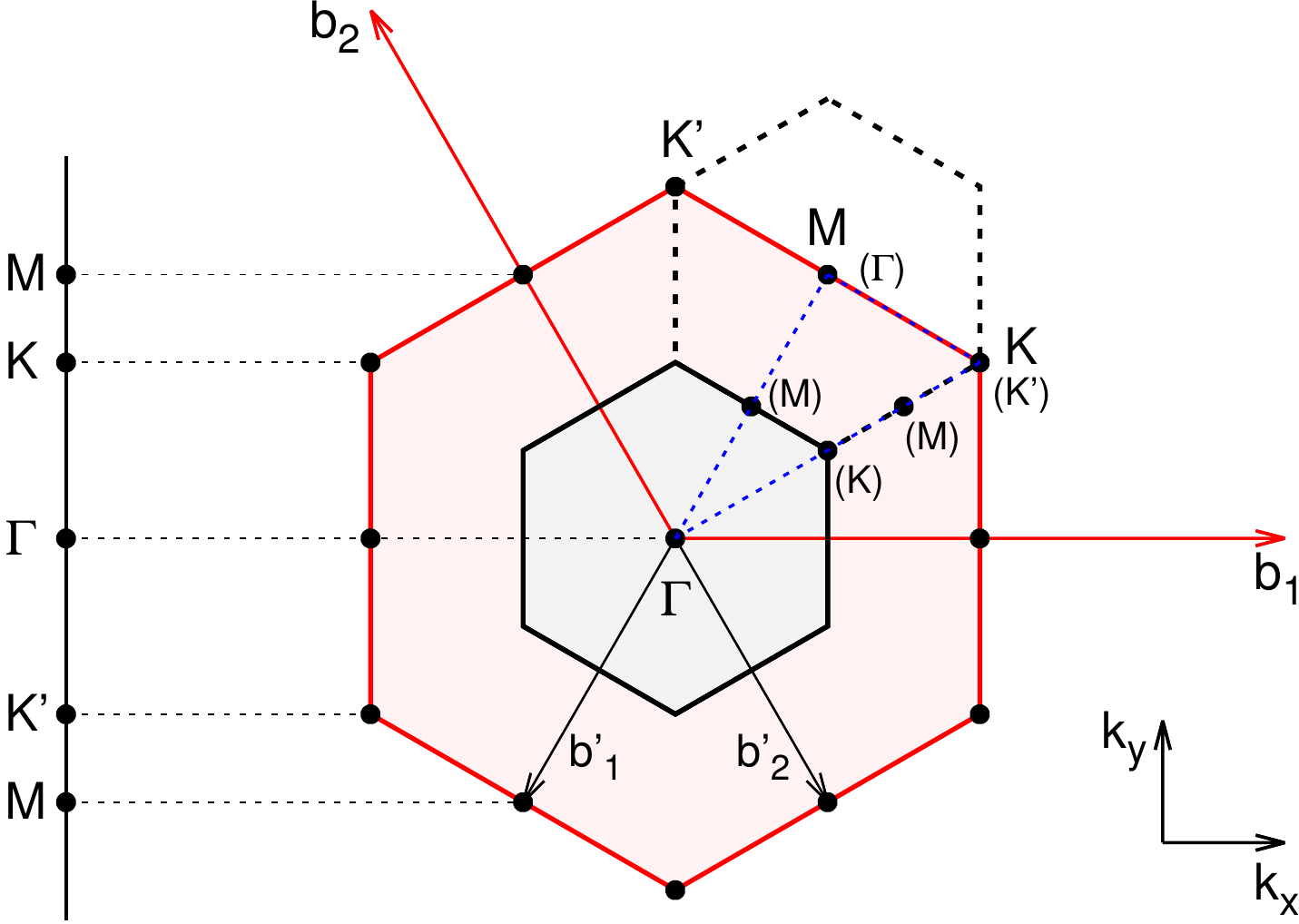}
    \caption{(Colour online.) Geometric Brillouin zone and effective Jones zone for the $(p,q)=(1,10)$ twist bilayer commensuration (6.6$^\circ$ rotation angle). The effective Jones zone is governed by the fundamental momentum scale at which states are coupled by the interlayer interaction, $g^{(c)} = \frac{8\pi}{\sqrt{3}a}\sin\frac{\theta}{2}$, and depends only on the twist angle, while the geometric Brillouin zone depends on the particular commensuration. The geometric zone is electronically equivalent as it entails simply a back-folding of states from the Jones zone without coupling, i.e., an irrelevant relabeling of momentum. The path illustrated in the figure is that which the band structure is plotted on in Fig.~\ref{Mgapbnd}.}
    \label{BZJones}
\end{figure}

\begin{figure}[t!]
    \includegraphics[width=0.9\linewidth]{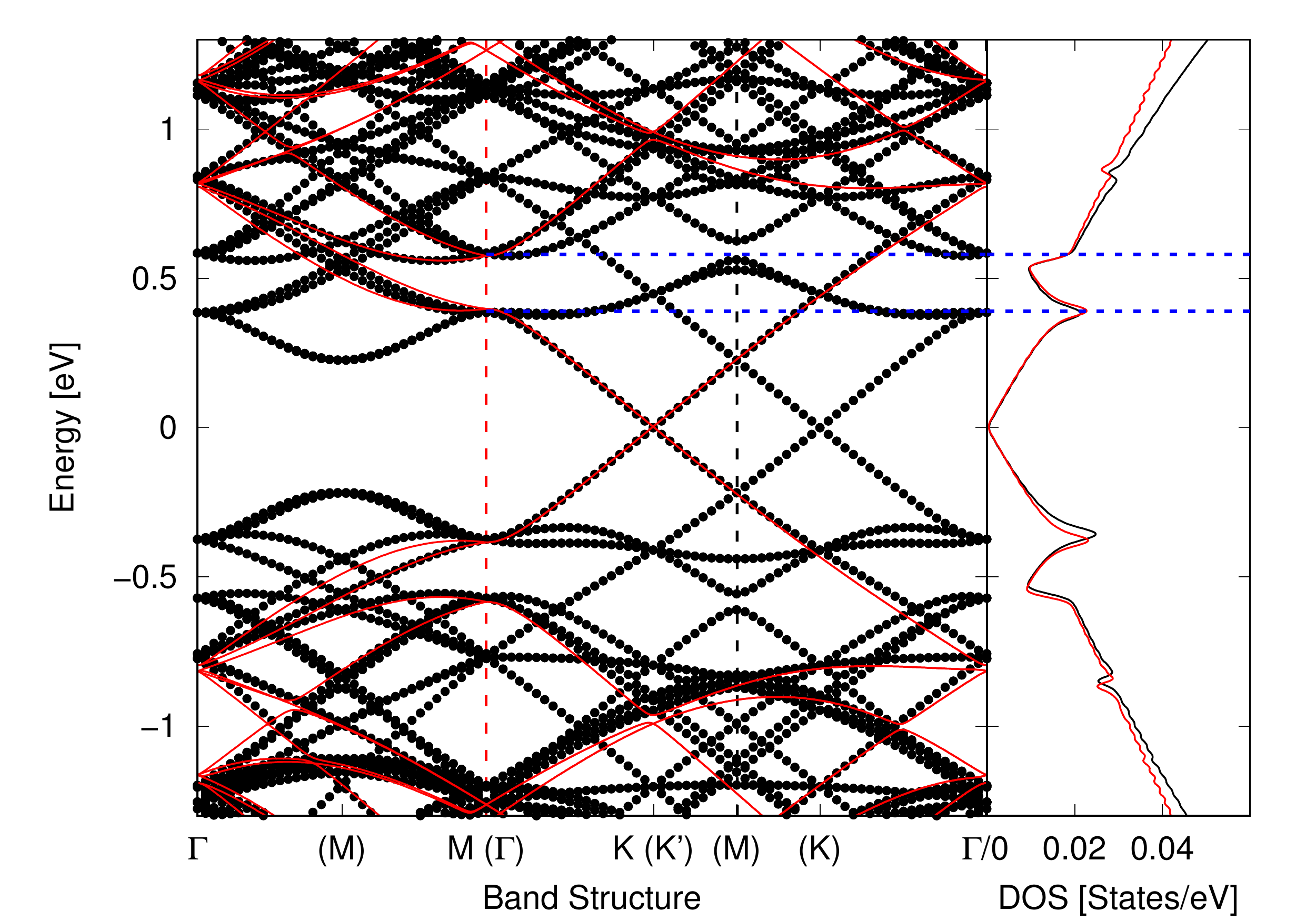}
    \caption{(Colour online.) Band structures from tight-binding -- light shaded (red) points -- and the effective Hamiltonian (black lines), with the right hand side panel exhibiting the corresponding density of states. On the band structure axis the larger symbols refer to the high symmetry points in the Jones zone, with the smaller symbols in parenthesis the high symmetry points of the geometric Brillouin zone (the band structure path is illustrated in Fig.~\ref{BZJones}.}
    \label{Mgapbnd}
\end{figure}

We first recall that for the bulk of the twist bilayer, as has been shown in Ref.~\onlinecite{shall13}, electronic properties depend solely on the twist angle, a fact that follows as all matrix elements of the interlayer interaction couple only on a momentum scale given by $g^{(c)} = \frac{8\pi}{\sqrt{3}a}\sin\frac{\theta}{2}$. For the bulk of the twist bilayer, therefore, the (commensuration dependent) geometric Brillouin zone is physically irrelevant and all physics is governed by an angle dependent and interaction governed Jones zone. Different choices of commensuration for the same twist angle then simply amount to an irrelevant relabeling of momentum in which states are back-folded, without interacting, into are smaller geometric Brillouin zone from the large Jones zone. Such relabeling will, however, play a non-trivial role in the \emph{projection} of the bulk band structure onto the 1d BZ of the nanoribbon, as all bulk states that project to the same k-vector of the 1d zone can, in principle, be coupled by scattering at the edge. The edge physics of the twist bilayer is therefore expected to be sensitive to details of the twist commensuration in a way not found in the bulk.

To see this in Fig.~\ref{rbnd}(c,d) we present the band structure of a nanoribbon with a $(p,q)=(1,10)$ commensuration that possess a twist angle $\theta = 6.6^\circ$, close to that found for the $(p,q)=(1,11)$ system ($\theta = 6.01^\circ$). Comparison with Fig.~\ref{rbnd}(a,b) shows that the moir\'e edge state is now no longer seen, although the zero mode is still found, and indeed is significantly increased in prominence as this commensuration generates edges dominated by the zigzag termination. It may also be noted that the $M$ point gap has vanished, and one might therefore suspect that the survival of the moir\'e edge state is connected to the existence of this gap. We will now investigate this question in detail.

\subsection{The $M$ point gap: physics beyond the twist angle}

For the $M$ point gap to exist in the 1d Brillouin zone of the bilayer, an obvious minimal requirement is that it is found in the bulk BZ. We will therefore first address the question of the existence of the $M$ point gap in the bulk. Interestingly, this gap -- which arises from the lowest energy intersection of the two Dirac cones and their consequent band repulsion -- is found only for certain twist commensurations, a fact noticed, but not explained, in Ref.~\onlinecite{jia13}.

As the bulk band structure is essentially independent of the details of lattice commensuration such a situation can only arise from the back folding of states from the underlying Jones zone, whose geometry is governed by the coupling momentum scale $g^{(c)}$, to the lattice Brillouin zone. In short, whether an $M$ point gap is seen or not simply depends whether the $M$ point of the Jones zone back-folds to the $M$ point of the geometry Brillouin zone, or to some other point. Purely geometric considerations determine this, and we find that and $M$ to $M$ point back folding occurs when the condition $\gamma = 6/\delta$ is satisfied (i.e., for exactly 50\% of possible commensurations).

To demonstrate this explicitly we calculate the electronic structure for a $(p,q)=(1,10)$ commensuration for which the Jones and Brillouin zones are not equal: $A_{\text{Jones}} = 4A_{\text{BZ}}$. For this we must use two different methods. The tight-binding method can be employed to generate the band structure for the exact $(p,q)=(1,10)$ commensuration, however to calculate the electronic structure in the Jones zone we must use an effective Hamiltonian approach that couples states only on the fundamental momentum scale $g^{(c)}=\frac{8\pi}{\sqrt{3}a}\sin\frac{\theta}{2}$. 

In Fig.~\ref{Mgapbnd} we show both the geometry of these two zones (panel (a)) along with the band structures plotted in the $\Gamma$-M-K-$\Gamma$ path of the larger Jones zone (panel (b)), with the corresponding special points of the smaller lattice BZ presented in parenthesis. As can be seen while there is an $M$ point gap in the physically relevant Jones zone, this is back-folded to the $\Gamma$ point of the lattice Brillouin zone, and hence at the $M$ point of this latter zone the two Dirac bands from each layer intersect without gap formation. Note that the density of states from both the tight-binding method and effective continuum Hamiltonian are, at low energies, in near perfect agreement, testifying to the correctness of the effective Hamiltonian presented in Ref.~\onlinecite{shall16}.

\begin{figure}[t!]
  \hspace{-1cm}
  \includegraphics[width=0.8\linewidth]{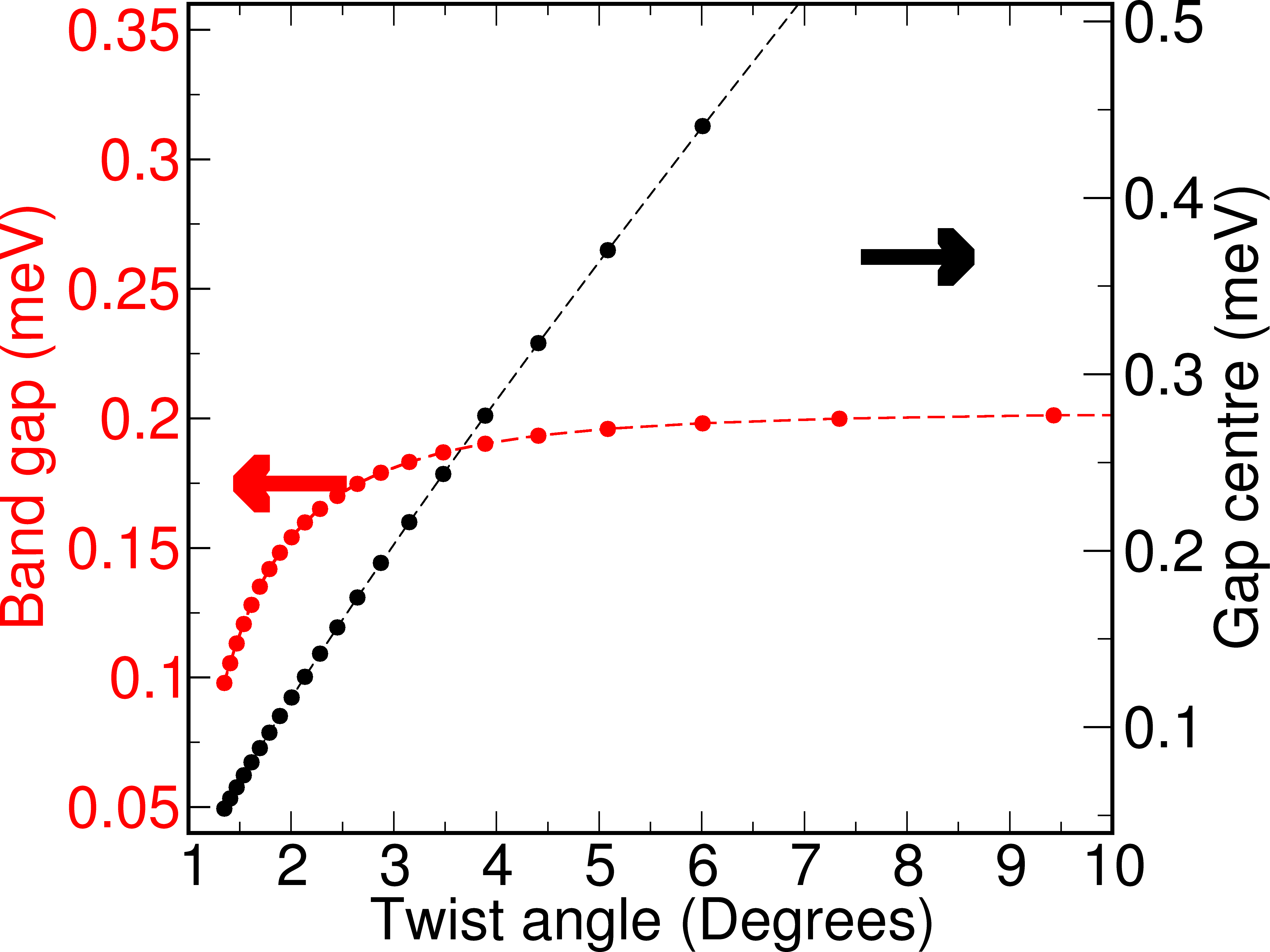}
  \caption{(Colour online.) Magnitude and energy centre of the $M$ point gap as a function of twist angle.}
  \label{M}
\end{figure}

As the $M$ point gap is evidently crucial for the existence of the moir\'e edge state, it is interesting to characterize the gap centre energy and gap size as a function of angle. As may be seen in Fig.~\ref{M}, the $M$ point gap is equal to $\sim\-\-0.2$~eV for $\theta > 5^\circ$ and falls to zero only in the small angle limit, behaviour which follows closely the form of the Fermi velocity. This should be contrasted with the very small Dirac point gap found only at large angles for the smallest twist bilayer unit cells.

\section{Conclusions}

The graphene twist bilayer has been investigated in a confining nanoribbon geometry, whose edges feature low index facets that are either armchair or zigzag dominated or, in the case of edge roughening, an approximately equal mixture of these two terminations. Irrespective of the atomic structure, we find these edges are associated with novel ``moir\'e edge states'' whose energies coincide with the energies of van Hove singularities occurring in the bulk electronic structure, and that possess an electron density strongly modulated by the moir\'e lattice. These edge states are therefore both robust against edge roughness as well as possessing an energy tunable by the rotation angle, properties that stand in contrast to the zigzag zero mode for which the sensitively to roughness has made experimental observation very difficult\cite{wang16}. Such moir\'e edge states thus represent an interesting alternative to the zero mode  for exploring the wealth of physical phenomena predicted for edge states in graphene based systems. Future questions of interest are to examine the possibility of a topological origin of moir\'e edge states (given their insensitivity to boundary conditions), as well as their relation to the localized states found at partial dislocations in bilayer graphene\cite{butz13,shall17,yin16}.

\begin{acknowledgments}

This work was supported by the Collaborative Research Center SFB 953 of the Deutsche Forschungsgemeinschaft (DFG).

\end{acknowledgments}



\end{document}